\documentclass[aps,pre,twocolumn]{revtex4-1}
\usepackage{amsfonts,amssymb,amsmath,bm}
\usepackage{graphicx,epsfig,color,bm}

\begin{document}

\title{Scale-dependent co-localization in a population of gyrotactic swimmers}
\author{M. Borgnino$^1$, F. De Lillo$^1$ and G. Boffetta$^1$}

\affiliation{
$^1$Department of Physics and INFN, Universit\`a di Torino,
via P. Giuria 1, 10125 Torino, Italy 
}

\date{\today}

\begin{abstract}
We study the small scale clustering of gyrotactic swimmers transported by a turbulent flow, when the intrinsic variability of the swimming parameters within the population is considered.
By means of extensive numerical simulations, 
we find that the variety of the population introduces
a characteristic scale $R^*$ in its spatial distribution. At scales smaller than
$R^*$ the swimmers are homogeneously distributed, while at larger scales 
an inhomogeneous distribution is observed with a fractal dimension
close to what observed for a monodisperse population characterized 
by mean parameters.
The scale $R^*$ depends on the dispersion of the population and it is 
found to scale linearly with the standard deviation both for a 
bimodal and for a Gaussian distribution.
Our numerical results, which extend recent findings for a 
monodisperse population, indicate that in principle it is possible 
to observe small scale, fractal clustering in a laboratory experiment
with gyrotactic cells.
\end{abstract}


\maketitle

\section{Introduction}
\label{sec1}
The microscopic distribution of aquatic microorganisms has profound
effects on the ecology of the oceans \cite{Williams2011,Mitchell2008}. One example is the observed patchiness 
of phytoplankton at the sub-meter scale which has a fundamental impact on the 
rate at which cells encounter each other and their predators\cite{Kiorboe2008,Visser2006}.
Patchiness of phytoplankton at different scales has different origins. While 
at large scales it is driven by reproduction and/or nutrients\cite{Mackas1985,Martin2003}, 
at scales smaller than one
kilometer patchiness is expected to be produced by physical mechanisms,
including plankton motility and the interaction with the flow .
Indeed, field observations have revealed that motile phytoplankton are 
considerably more patchy at small scales than non-motile species 
\cite{mouritsen2003vertical,Malkiel1999}.

Several species of motile phytoplankton are able to swim in the 
vertical direction guided by a stabilizing torque arising from an unbalance 
distribution of the mass in the cell \cite{Pedley1987,durham2009disruption}. 
The resulting swimming direction of these 
gyrotactic cells stems from the competition between the 
stabilizing torque and the shear-induced viscous torque \cite{kessler1985hydrodynamic,Thorn2010,OMalley2012,Lewis2003}.
Numerical and experimental works have revealed how gyrotactic motility, combined with the presence of a flow, generates strongly inhomogeneous distributions.
In the case of laminar flow, gyrotaxis produces a beam-like accumulation
in downwelling pipe flows \cite{kessler1985hydrodynamic}, while in horizontal
shear flow it generates accumulation in thin layers 
\cite{durham2009disruption,Durham2012,Santamaria2014}.
Recent works have shown that gyrotaxis also produces clustering at very
small scales (comparable with the Kolmogorov scale) in non-stationary
turbulent flows\cite{durham2013turbulence,de2014turbulent,Zhan2013,gustavsson2016preferential}. In this case cells are found to accumulate on fractal
dynamical clusters characterized by a fractal dimension which depends on the
cell and flow parameters 
\cite{durham2013turbulence,de2014turbulent,Gustavsson2016}.

In this work we consider the dynamics of an inhomogeneous population of 
gyrotactic cells, characterized by a distribution of cells' parameters,
transported by a turbulent flow. 
The motivation of our study is to determine 
the robustness of fractal clustering induced by turbulence
on a distribution of cells with slightly different biological parameters,
typical of a natural population. 
The main result, obtained by means of extensive numerical simulations,
is that fractal clustering is observable, at large enough scales, also in populations with
significant variability (up to $20 \, \%$ of relative variation in 
gyrotactic parameters).
Moreover, by considering a simplified bimodal population, we introduce
a crossover scale (above which fractal clustering is observable) and we
predict how this scale depends on the population variability.

The remaining part of this paper is organized as follows. In Section \ref{sec2}
we introduce the mathematical model for gyrotactic swimmers and we discuss,
on the basis of simple arguments, how clustering depends on the 
population distribution. Section \ref{sec3} is devoted to numerical results
for two particular distributions, while section \ref{sec4} summarizes
our results.

\section{Mathematical model}
\label{sec2}
We consider the classical model of gyrotactic swimmers which describes the
motion of a bottom-heavy spherical cell \cite{kessler1985hydrodynamic,Pedley1992}
at position ${\bm X}$ swimming in the direction ${\bm p}$
(with $|{\bm p}|=1$)
\begin{eqnarray}
{d {\bm X} \over dt} &=& {\bm u}({\bm X},t) + V {\bm p}\,,
\label{eq:1} \\
{d {\bm p} \over dt} &=& {1 \over 2 B}\left[{\bm k} - ({\bm k} \cdot {\bm p}) {\bm p}\right]
+{1 \over 2} {\bm \omega}({\bm X},t) \times {\bm p}\,
\label{eq:2}
\end{eqnarray}
where ${\bm u}({\bm x},t)$ is the velocity field, ${\bm \omega}={\bm \nabla}
\times {\bm u}$ is the vorticity, ${\bm k}=(0,0,1)$ is the vertical unit vector.
The first term on the rhs of (\ref{eq:2}) represents the effect of the 
gravitational torque which orients the swimming direction towards the 
vertical, while the last term is viscous torque which rotates the cells
with the local vorticity.
$V$ is the swimming velocity, assumed constant, 
while $B=3 \nu/(g h)$ is the gyrotactic reorientation time where $\nu$ is the
kinematic viscosity of the fluid, $g$ the acceleration of gravity and $h$ 
measures the displacement of the center of mass from the geometrical
center of the cell.

The gyrotactic swimmers are transported by a turbulent velocity field
${\bm u}({\bm x},t)$ obtained by direct numerical simulations (DNS) 
of the incompressible Navier-Stokes equations
\begin{equation}
\partial_t {\bm u} + {\bm u} \cdot
{\bm \nabla} {\bm u} = - {\bm \nabla} p + \nu \nabla^2 {\bm u} + {\bm f}
\label{eq:3}
\end{equation}
where ${\bm f}$ represents a zero-mean, temporally uncorrelated Gaussian 
forcing which injects energy at large scales at a rate $\varepsilon$.
Together with the viscosity, the energy injection rate defines the 
Kolmogorov length scale $\eta_{K}=(\nu^3/\varepsilon)^{1/4}$, the
Kolmogorov time scale $\tau_{K}=(\nu/\varepsilon)^{1/2}$ and the
Kolmogorov velocity $v_{K}=\eta_K/\tau_K=(\nu \varepsilon)^{1/4}$
\cite{frisch1995turbulence}.
These characteristic scales are used to make the parameters
in the gyrotactic model dimensionless. The ratio of the two terms on the rhs of (\ref{eq:1})
defines the swimming number $\Phi \equiv V/v_{K}$, while the ratio of 
the two terms in (\ref{eq:2}) gives the stability number 
$\Psi \equiv B/\tau_{K}$.

Formally, equations (\ref{eq:1}) and (\ref{eq:2}) define a dissipative dynamical
system in the $({\bm X},{\bm p})$ phase space of dimension $2d-1$ ($d=3$) 
with an expansion rate in the phase space given by
\begin{equation}
\sum_{i=1}^{d} \left({\partial \dot{X}_i \over \partial X_i} 
+ {\partial \dot{p}_i \over \partial p_i} \right) = - {d-1 \over 2 B} p_3 \, .
\label{eq:4}
\end{equation}
As the swimming direction orients towards the vertical direction
($p_3>0$)
the expansion rate becomes negative and the trajectories collapse 
on a fractal attractor in the phase space. When the attractor has 
dimension less than $d$ the swimmers concentrate (in physical space) on clusters
with the same fractal dimension 
\cite{paladin1987anomalous}.

When the swimming number vanishes (i.e. $V=0$) the cells in (\ref{eq:1})
are simply transported by an incompressible velocity field and therefore
they cannot accumulate (as (\ref{eq:1}) decouples from (\ref{eq:2})). 
Moreover, when $B$ is smaller than the Kolmogorov time, i.e. $\Psi \ll 1$,
we can expand (\ref{eq:2}) at the first order in $B/\tau_{K}$ to obtain,
in stationary conditions \cite{durham2013turbulence}
\begin{equation}
{\bm p} \approx (B \omega_y, -B \omega_x, 1)
\label{eq:5}
\end{equation}
which shows that when $B=0$, ${\bm p}$ is aligned
towards the vertical direction and the motion of the swimmers is given 
by the superposition
of an incompressible velocity and a uniform vertical migration which, again,
cannot produce clustering. 
Similarly, for $B \to \infty$ the expansion rate (\ref{eq:4}) vanishes and 
also in this case swimmers are not expected to cluster.
Previous numerical simulations have shown that
indeed gyrotactic swimmers produce clusters for intermediate values of 
$\Psi$ (and $\Phi>0$) with maximum clustering for $\Psi \simeq 1$
\cite{durham2013turbulence}.

To quantify the degree of clustering we use the correlation dimension
$D$, defined as the scaling exponent of the probability to find two
particles at a distance less than $r$: 
$P(|{\bm X}_1-{\bm X}_2|<r) \propto r^{D}$ as $r \to 0$ 
\cite{paladin1987anomalous}. For homogeneous distribution in space 
one has $D=d$, while $D<d$ indicates fractal clustering. 

When considering a population of swimmers with different parameters 
$V$ and $B$ we can extend the above definition to measure the 
{\it cross-correlation dimension} $D_{12}(r)$ defined in terms of the
probability of finding two swimmers characterized by two sets of
parameters $(V_1,B_1)$ and $(V_2,B_2)$ at a distance smaller than $r$:
$P_{12}(r) \propto r^{D_{12}}$ \cite{Bec2005}.
In principle, we cannot expect a power-law scaling for $P_{12}(r)$ for
a generic couple of swimmer parameters and therefore $D_{12}$ is a 
function of $r$ and not simply a scaling exponent. 
Of course, for a monodisperse population, with $V_2=V_1$ and $B_2=B_1$ the
cross-correlation dimension recovers the correlation dimension of the
population, $D_{12}(r)=D$.

Consider now a couple of swimmers at positions ${\bm X}_1$ and 
${\bm X}_2={\bm X}_1+{\bm R}$
with slightly different parameters, e.g.
with the same swimming velocity $V_2=V_1$ and with different reorientation
time $B_2=B_1+\Delta B$. We assume that $\Delta B$ is a small parameter
such that the separation between the two trajectories ${\bm R}$ is
smaller than the Kolmogorov scale. 
According to (\ref{eq:1}) this separation evolves according to
\begin{equation}
{d {\bm R} \over dt} = \Delta {\bm u}({\bm R}) + V \Delta {\bm p}
\label{eq:6}
\end{equation}
where $\Delta {\bm u}({\bm R})={\bm u}({\bm X}_2)-{\bm u}({\bm X}_1)$
and $\Delta {\bm p}={\bm p}_2-{\bm p}_1$.

The first term on the rhs of (\ref{eq:6}) 
is proportional to $v_{K} (R/\eta_{K}$),
while the second term, in the limit of small stability numbers, 
contains the difference $ \omega \Delta B$. The
ratio of these two terms defines a characteristic scale 
$R^{*} \simeq \eta_{K} \Phi \Delta \Psi$. 
When $R<R^{*}$ the swimmer velocity difference is dominated
by the second term in (\ref{eq:6}): the two trajectories are 
uncorrelated and one swimmer sees the other population as
uniformly distributed. On the contrary, when $R>R^{*}$, the first 
term in (\ref{eq:6}) dominates and the correlations between the
two population, induced by the common velocity field, 
appears \cite{bec2005clustering}.
Therefore, for a bimodal distribution, we expect two different 
behaviors for $D_{12}(r)$: $D_{12}(r)=3$ for $R<R^{*}$ and
$D_{12}(r) \simeq D_{11}$ for $R>R^{*}$ 
($D_{11} \simeq D_{22}$ are the correlation dimensions of the two 
populations, which are close by hypothesis).

In the case of two swimmers with the same reorientation time $B$ and
different swimming velocity $V_1$ and $V_2=V_1+\Delta V$, a similar
argument, in the limit of small parameter difference, leads again to 
a characteristic scale $R^{*} \simeq \eta_{K} \Psi \Delta \Phi$ which 
separates scales with homogeneous and fractal relative distribution.

The general case of a polydisperse population, characterized by a 
probability density function of parameters $f(V,B)$ is the most interesting
for applications to experimental data where one cannot avoid the natural 
intrinsic variability of the population. 
Also in this case we will consider the cumulative probability of having
two swimmers at a distance lower than $r$, integrated over the distribution
$f(V,B)$. Again, for very small $r$ we expect  
this probability to decrease proportional to $r^3$ as different cells in the 
population are spatially decorrelated. The interesting question is 
whether also for a continuous distribution of cell parameters 
with finite support there exists a characteristic scale $R^*$ above 
which a fractal dimension can be observed which can be interpreted as 
that of an "average", monodisperse population. 
To address this point we will consider a population characterized
by a Gaussian parameter distribution $f(B)$ with mean value $\bar{B}$ 
and variance $\sigma_B$.

\section{Numerical results}
\label{sec3}
We have performed a numerical investigation of the spatial 
distribution of several populations of swimmers, characterized by
different distributions $f(V,B)$ of swimming parameters. 
The velocity (and vorticity) field in (\ref{eq:1}-\ref{eq:2}) are 
obtained by a direct numerical simulations of the NS equations 
(\ref{eq:3}) by using a fully dealiased pseudo-spectral code at 
different resolutions.
After the flow has reached a statistical steady state, a population 
of $N_s$ cells is initialized with uniform random positions ${\bm X}$ 
in the domain and and orientation ${\bm p}$ on the unit sphere. 
The motion of the swimmers is obtained by the simultaneous
integration of (\ref{eq:3}) and (\ref{eq:1}-\ref{eq:2}) in which 
fluid velocity and 
vorticity at the cell positions are obtained by trilinear interpolation
\cite{biferale2004multifractal}.
After the swimmer distribution has reached a statistical steady state, 
we collect data for several large-scale eddy turnover times to ensure 
statistical convergence.

\begin{figure}[h!]
\includegraphics[width=\columnwidth]{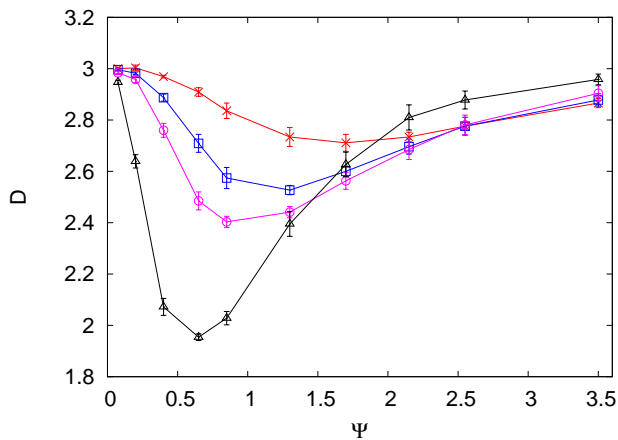}
\caption{(Color online). 
Correlation dimension $D$ for a homogeneous population of gyrotactic 
swimmers as a function of the stability number $\Psi$. 
Different lines correspond to different swimming numbers: 
$\Phi=0.33$ (red crosses), $\Phi=0.66$ (blue squares), $\Phi=1.0$
(purple circles) and $\Phi=3.0$ (black triangles). 
The error bars are estimated on the fluctuations of the dimension
with the statistics.}
\label{fig1}
\end{figure}

In Figure~\ref{fig1} we plot the correlation dimension for a monodisperse
population as a function of the swimming number $\Phi$ and stability
number $\Psi$. As already reported, clustering is maximum (i.e. $D$ is
minimum) for $\Psi \simeq 1$ and large $\Phi$
\cite{durham2013turbulence,gustavsson2016preferential}
while $D \simeq 3$ for both large and small values of $\Psi$, as 
discussed in Section~\ref{sec2}.
The position of the minimum $D$ (maximum clustering) 
depends on the swimming velocity as, for
small $\Psi$, one has $3-D \propto (\Phi \Psi)^2$ \cite{durham2013turbulence}.

\subsection{Bimodal distribution}
We first consider a bimodal population composed by two species with the same
swimming number $\Phi$ and different stability numbers
$\Psi_1$ and $\Psi_2=\Psi_1+\Delta \Psi$ with the same number of cells in each 
species, i.e. with marginal distribution
$f(B)={1 \over 2} \delta(B-B_1) + {1 \over 2} \delta(B-B_2)$.
The difference $\Delta \Psi$ defines the standard deviation of the 
distribution $\sigma_B=\Delta \Psi/2$.

\begin{figure}[h!]
\includegraphics[width=\columnwidth]{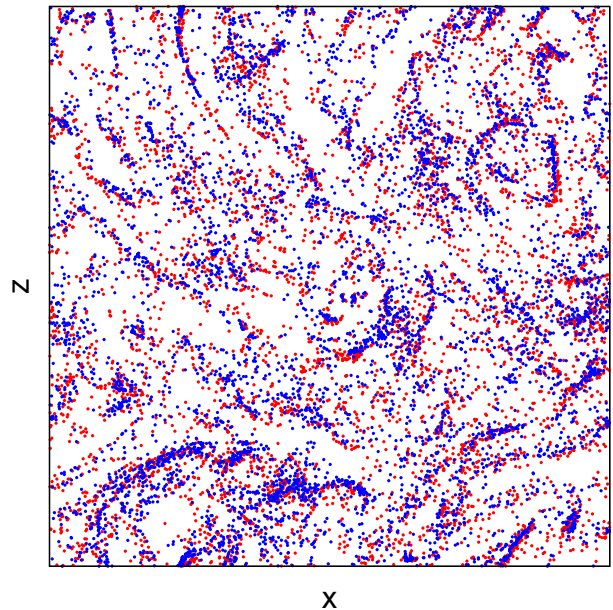}
\caption{(Color online). 
Vertical section of the positions of two species of swimmers with
$\Psi_1=0.5$ (red), $\Psi_2=0.667$ (blue) and $\Phi_1=\Phi_2=3.0$ in a 
turbulent flow.}
\label{fig2}
\end{figure}

Figure~\ref{fig2} shows a 2D section of the 3D distribution of a bimodal 
population with dimensionless parameters $\Psi_1=0.5$, $\Psi_2=0.667$ and 
$\Phi_1=\Phi_2=3.0$ with the two species plotted with 
different color.
Both species are expected to cluster according to the 
results shown in Fig.~\ref{fig1} with correlation dimension $D \simeq 2.0$.
It is evident that at large scales the distributions of the two populations 
display similar features, 
while at small scales different distributions appear,
in qualitative agreement with the argument discussed in Section~\ref{sec2}.

\begin{figure}[h!]
\includegraphics[width=\columnwidth]{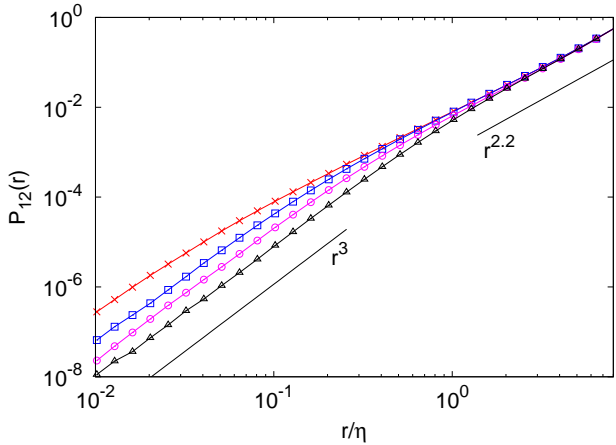}
\caption{(Color online). 
Probability $P_{12}(r)$ to find two cells of different populations $1$ and $2$
at distance smaller than $r$ for different pairs of population parameters with
$\Delta \Psi=0.0042$ (red crosses), $\Delta \Psi=0.021$ (blue squares),
$\Delta \Psi=0.042$ (purple circles) and $\Delta \Psi=0.125$ (black triangles).
Each population is composed by $6.4 \times 10^4$ individuals.}
\label{fig3}
\end{figure}

The scale-dependent co-localization is quantified by the cross probability
$P_{12}(r)$ plotted in Fig.~\ref{fig3} for pairs of populations
with different values of 
$\Delta \Psi$. We see that, for all pairs considered, the probability 
displays a scaling close to $r^3$ at very small scales confirming that, 
at these scales, the two populations have uncorrelated distributions. 
On the contrary, for sufficiently large scales, the probability distribution
follows a power-law scaling with exponent $\simeq 2.2$, close to the 
fractal dimension of a homogeneous population with stability 
number $\overline{\Psi}$, the average of the two species.

\begin{figure}[h!]
\includegraphics[width=\columnwidth]{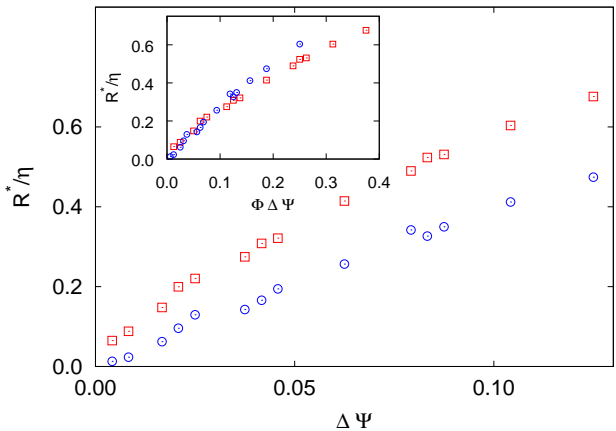}
\caption{(Color online). 
Crossover scale $R^{*}$ as a function of $\Delta \Psi$ for two
different set of populations with $\Phi=3$ (red crosses) and 
$\Phi=1.5$ (blue squares). 
Inset: the same data plotted as a function of $\Phi \Delta \Psi$.}
\label{fig4}
\end{figure}

The transition between the two scaling ranges, although broad, clearly
moves to larger scale as the difference $\Delta \Psi$ increases.
In order to quantify this transition, we computed the crossover scale
$R^{*}$ defined empirically by the intersection of two power-law fits
of $P_{12}(r)$ at small scales and large scales respectively.
The small scale exponent is close to $3$ (we obtain an exponent between
$2.8$ and $3.0$ for all the case considered), while the large scale exponent
depends on $\overline{\Psi}$.
Figure~\ref{fig4} shows the dependence of $R^*$ on the population
variance $\Delta \Psi$, for different swimming number $\Phi$,
which confirms the linear scaling of $R^*$ predicted in Section~\ref{sec2}.
The inset of Fig.~\ref{fig4} shows the remarkable collapse of $R^*$
when plotted as a function of $\Phi \Delta \Psi$,
as predicted in Section~\ref{sec2}.

\begin{figure}[h!]
\includegraphics[width=\columnwidth]{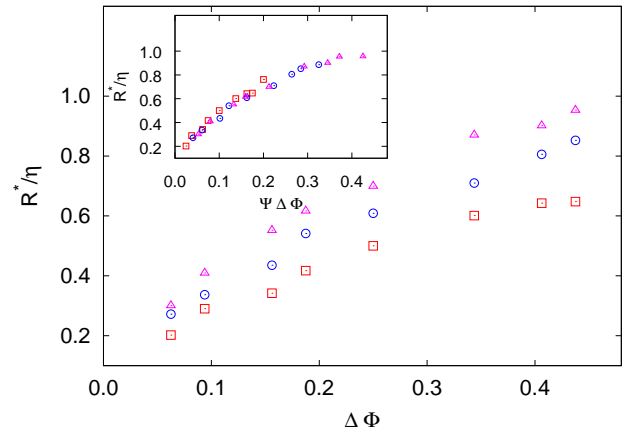}
\caption{(Color online). 
Crossover scale $R^{*}$ as a function of $\Delta \Psi$ for two
different set of populations with $\Psi=0.4$ (red crosses),
$\Psi=0.65$ (blue squares) and $\Psi=0.85$ (pink triangles).
Inset: the same data plotted as a function of $\Psi \Delta \Phi$.}
\label{fig5}
\end{figure}

A similar behavior is observed when considering a bimodal population with
two different swimming numbers $\Phi_1$ and $\Phi_2=\Phi_1+\Delta \Phi$.
Figure~\ref{fig5} refers to three examples of bimodal populations 
characterized by three different stability numbers $\Psi$ close to the 
value for maximum clustering shown in Fig.~\ref{fig1}
($\Psi=0.4$, $\Psi=0.65$ and $\Psi=0.85$). 
Also in this case, different scaling behaviors of $P_{12}(r)$ are observed for 
small and large separations and the fit of these scaling laws are used 
to define the crossover scale $R^*$ plotted in figure.
The inset of Fig.~\ref{fig5} shows a good collapse of the different
crossover scales when plotted as a function of the combination 
$\Phi \Delta \Psi$, confirming that this is the relevant parameter in
the process. 

\subsection{Gaussian distribution}
We now consider the more realistic case of a population of swimmers
with stability number $\Psi$ following a Gaussian distribution with mean
value 
$\overline{\Psi}$ and standard deviation $\sigma_\Psi$. Having in mind an
experimental study in which we do not know the value of $\Psi$ (i.e. $B$)
for each individual, we consider the cumulative probability $P(r)$ of 
having two cells at a distance smaller than $r$ integrated on all the 
possible pairs in the population. 

\begin{figure}[h!]
\includegraphics[width=\columnwidth]{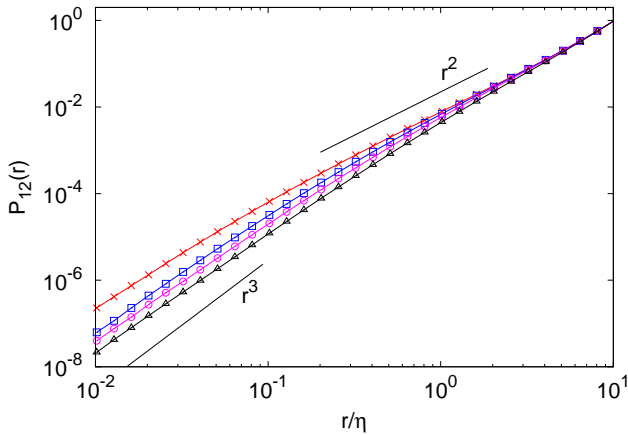}
\caption{(Color online). 
Probability $P(r)$ to find two cells at distance smaller than $r$ for 
a population of gyrotactic swimmers with fixed $\Phi=3$ and $\Psi$ Gaussian 
distributed with $\overline{\Psi}=0.583$ and $\sigma_\Psi=0.008$ 
(red crosses), $\sigma_\Psi=0.042$ (blue squares),
$\sigma_\Psi=0.083$ (purple circles) and $\sigma_\Psi=0.166$ (black triangles).
Each population is composed by $3 \times 10^5$ individuals.
Inset: Crossover scale $R^*$ as a function of $\sigma_{\Psi}$.}
\label{fig6}
\end{figure}

The dependence of $P(r)$ on $r$ is shown in Fig.~\ref{fig6} for several
populations with different standard deviations 
$\sigma_{\Psi}$. Similarly to the case 
of bimodal distribution, we recognize different ranges of scales.
At very small scale, $P(r)$ converges towards the uniform scaling $r^3$,
more clearly for the case with larger variance while for small variance 
the a smaller scaling exponent is observed (between $2.6$ and $2.8$).
At larger scales, $r \gtrsim \eta$, we observe a different power-law 
behavior with an exponent which weakly depends on $\sigma_{\Psi}$ and is 
very close to the exponent of a monodisperse population 
$D(\overline{\Psi})\simeq 2$ for the smallest variance while grows to 
above $2.3$ for the population with largest varince.
As in the case of bimodal distribution, also in this case the transition
from homogeneous ($D \simeq =3$) to fractal ($D \simeq 2$) distribution
moves to larger scales as $\sigma_{\Psi}$ increases, as shown in 
Fig.~\ref{fig6}. 
It is remarkable that also for the largest standard deviation, for which
$\sigma_{\Psi}/\overline{\Psi} \simeq 0.29$, the distribution of the population
at large scales is strongly inhomogeneous and the probability $P(r)$ 
indicates a fractal dimension close to $D(\overline{\Psi})$.

\section{Conclusions}
\label{sec4}

We have studied, by means of direct numerical simulations, 
the small scale clustering of a population of
gyrotactic cells, characterized by a distribution of gyrotactic
parameters, swimming in a turbulent environment. 
The main goal of our work was to extend the results obtained for a 
monodisperse population to a more realistic population, characterized
by a distribution of the swimming parameters. 


We considered two very different families of test populations: bimodal
populations, made of two hypothetical strains with different swimming 
or stability number,
and a more realistic case where the swimming number is Gaussian-distributed
within the population. 
Despite the differences between the distributions considered,
they show similar features for what concerns small scale clustering.
In all cases, the probability of finding inter-particle distances less than
$r$ exhibits two scaling ranges $r^{3}$ and $r^{D}$ for separations smaller and
larger than  a crossover scale $R^*$, respectively. 
The exponent $D$ represents the effective correlation dimension of 
the distribution when it is coarse-grained at a scale $R^*$. 
The crossover scale grows with the variance of the distribution,
confirming the linear dependence predicted for a narrow bimodal 
distribution.
Furthermore, in this case our numerical data confirm the prediction 
that $R^*$ depends on the product of the two dimensionless 
swimming parameters.

From an experimental point of view, our results allows one to estimate
a-priori, based on biological and fluid-dynamical data, whether clustering is
expected in a given range of scales for a given species. This should be taken
into account in designing or analyzing field measurements in relation to
turbulence-induced phytoplankton patchiness. Of course, analogous
considerations apply every time fractal clustering is predicted, with a fractal
dimension depending on parameters with a non-negligible intrinsic variability,
as exemplified by works on inertial-particle transport in turbulence
\cite{bec2005clustering}.

Our findings should help assess the ecological relevance of turbulence-induced
demixing \cite{durham2013turbulence}. Fractal clustering implies smaller
distances between neighboring cells with respect to a homogeneous distribution
with the same average density. This has consequences for mating, resource
exploitation and risk of predation. Consideration of the variability in
swimming parameters might lead one to conclude that small scale clustering is
in practice irrelevant. However, if indeed the distribution is fractal on a
finite range of scales, the effect on nearest-neighbor-distance could be
diminished but still relevant. If predation by zooplankton is considered, $R^*$
could be larger than the typical perception radius of the predator (e.g. a
copepod or a fish larva), which would detect a locally homogeneous distribution
of prey, but smaller than the typical swimming distances covered while cruising
for prey, so that the underlying fractality might still have consequences
for the predation strategy \cite{sims2008}. Moreover, the possibility of a heterogeneous population to retain a fractal distribution on larger scales may have effects for population dynamics \cite{Benzi2012, Pigolotti2012}.
\begin{acknowledgments}
This article is based upon work from COST Action MP1305, supported by COST
(European Cooperation in Science and Technology).
We thank M. Cencini for fruitful discussion. 
R. Stocker is acknowledged for hospitality and useful discussions.
\end{acknowledgments}

\bibliography{biblio}

\end{document}